\documentstyle[12pt,A4]{article}%	for bibname
\date{11-Apr-1997}%	added paragraph on invariant lagrangians
\makeatletter%			for uttermost LaTeX--compatib.
\def\warnred#1#2{\@ifundefined{#1}{}{\@warning{REDEFINING \expandafter\string
\csname #1\endcsname\space TO \string#2}}}
\warnred{nc}{newcommand}\let\nc\newcommand
\nc\mycomment[1]{\marginpar{\vspace*{-1.5\baselineskip}%	WARUM???
\tiny\begin{flushleft}#1\end{flushleft}}}
\nc\remark[1]{}%					to ignore a block
\nc\com{~,\hs}
\nc\hs{\hspace*{2em}}
\nc\mypar{\par\vskip 1ex}%			PARAGRAPH WITH SOME EXTRA SKIP
\nc\noi{\noindent}
\nc\xvspace{\par~\par}
\nc\hr{\xvspace\hrule\xvspace}
\nc\ms[1]{\rule{0pt}{#1}}%			MyStrut
\nc\SR[2]{\rule[#1]{0pt}{#2}}
\nc\vs[1]{\SR{-#1 mm}{#1 mm}}
\nc\vt[1]{\rule{0mm}{#1mm}}
\nc\vu[1]{\SR{#1 mm}{#1 mm}}
\nc\vv[1]{\SR{#1 mm}{1 mm}}
\nc\initeqn{%\setcounter{equation}0
\def\theequation{\thesection.\arabic{equation}}\@addtoreset{equation}{section}}
\nc\sect{\mynewpage\initeqn\section}%	(\initeqn should be called only once...)
\nc\mynewpage{{\newpage\thispagestyle{empty}\cleardoublepage
\thispagestyle{plain}}}%
\warnred{ul}{underbar}
\nc\bea{\begin{eqnarray}}
\nc\be{\begin{equation}}%	10=space,4=tab
\nc\bex{\begin{equation}\let\\=\com\catcode`\&=10\relax}%	10=space,4=tab
\nc\ee{\end{equation}}
\nc\ena{\end{eqnarray}}
\nc\bit{\begin{itemize}}
\nc\eit{\end{itemize}}
\nc\ba{\begin{array}}
\nc\ea{\end{array}}
\nc\frenchdefs[9]{
\hyphenation{ana-ly-tique ana-ly-se ana-ly-ser ca-no-nique ca-no-niques
coup-lage coup-lages de-scrip-tion diffe-rent diffe-rents diffe-rente
diffe-rentes en-cou-rage-ment en-cou-rage en-cou-rage-ments exemple
exo-tique exo-tiques fer-mio-ni-que fer-mio-ni-ques in-va-riant
in-va-riante jeune jeunes mo-di-fie mo-di-fies mo-di-fiees
in-va-riantes Karls-ruhe Le-gen-dre par-ti-cule par-ti-cules
re-nor-mali-sable re-nor-mali-sables super-espace super-champs
syme-trie syme-tries super-gravite va-ria-tion-nel va-ria-tion-nels
va-ria-tion-nelle va-ria-tion-nelles veri-fier veri-fie veri-fiant
veri-fient}
\nc\apparai[1]{appara{\^\i}#1}
\nc\cad{c'est-\`a-dire}
\nc\coder[1]{d\'e\-ri\-v\'ee#1 co\-va\-ri\-an\-te#1}%	ARG: {} or {s}
\nc\cotrans[1]{\susytrans{#1} covariante#1}%		ARG: {} or {s}
\nc\diff{diff\'e\-ren\-tiel}
\nc\diffop[1]{op\'erateur#1 \diff{#1}}%			ARG: {} or {s}
\nc\diffgeo{g\'eo\-m\'e\-trie \diff le}
\nc\eqref[1]{{\mbox{\'eq. (\protect\ref{#1})}}}
\nc\eqn{\'equation}%					ARG: {} or {s}
\nc\bewgl[1]{\eqn{#1} de mouvement}%			ARG: {} or {s}
\nc\idb[1]{identit\'e#1 de Bianchi}
\nc\MS{Mo\-d\`ele Standard}
\nc\MSS{\MS\ Su\-per\-sy\-m\'e\-tri\-que}
\nc\puisque{\hs\mbox{puisque }}
\nc\scc{super\-champ chiral}
\nc\sccs{super\-champs chiraux}
\nc\susy{super\-sym\'e\-tri}%			followed by: e, que...
\nc\susyalg{alg\`ebre de \susy e}
\nc\susytrans[1]{trans\-for\-ma\-tion#1 de \susy e}%	ARG: {} or {s}
\nc\vev[1]{va\-leur#1 moyenne#1 dans le vide}%		ARG: {} or {s}
\nc\wfaktor{{\kappa}}
\nc\mysect[3]{\remark{\newpage~\vfill}\sect{#1}\remark{\vfill{\Huge$$ #2 $$}\vfill}#3\remark{\newpage}}
\warnred{@oldhat}{\string\^ --- This is bad !!!}\let\@oldhat=\^
\@warning{REDEFINING \string\^ FOR \string\^i to work (@oldhat=old def)}
\def\^{\@ifnextchar i{\@oldhat\i\@gobble}{\@oldhat}}
}%	-----	END OF FRENCHDEFS	-----
\nc\savebot[1]{\addtolength\textheight{#1}}
\nc\savetop[1]{\savebot{#1}\advance\voffset-#1}
\nc\wider[1]{\advance\hoffset -#1 \advance\textwidth #1 \advance\textwidth #1}
\nc\mydraft{\def\mynewpage{\vfill\pagebreak}
	\input fullpage.sty \myheadings \savebot{2cm}}
\nc\mytwocol{\advance\columnsep 1em\twocolumn\parindent0pt}
\@ifundefined{greektex}{}{\endinput}
\newcommand\greeknewcommand[2]{\def#1{{#2}}}
\greeknewcommand{\al}{{\alpha}}
\greeknewcommand{\gm}{{\gamma}}
\greeknewcommand{\dt}{{\delta}}
\def\eps{{\epsilon}}
\greeknewcommand\veps{\varepsilon}%				MY PREFERENCE
\greeknewcommand{\la}{{\lambda}}

\greeknewcommand\vk{\varkappa}
\def\vp{\varphi}
\greeknewcommand{\th}{\theta}
\greeknewcommand{\om}{{\omega}}
\greeknewcommand{\ze}{\zeta}

\greeknewcommand{\Gm}{\Gamma}
\greeknewcommand{\Si}{\Sigma}
\greeknewcommand{\Dt}{\Delta}
\greeknewcommand{\La}{\Lambda}
\greeknewcommand{\Th}{\Theta}
\greeknewcommand{\Om}{{\Omega}}

\greeknewcommand{\bze}{\bar\zeta}
\greeknewcommand{\thb}{\bar\theta}
\greeknewcommand{\vpb}{\bar\vp}%		AKA \bvp,\bvar,...
\greeknewcommand\bvp{\bar\varphi}%
\greeknewcommand{\lab}{\bar\lambda}
\greeknewcommand{\xib}{\bar\xi}
\greeknewcommand{\psib}{\bar\psi}
\greeknewcommand{\chib}{\bar\chi}
\greeknewcommand{\phib}{\bar\phi}
\greeknewcommand{\zeb}{\bar\zeta}

\greeknewcommand{\Phib}{\bar\Phi}
\greeknewcommand{\Psib}{\bar\Psi}
\greeknewcommand{\Thb}{\bar\Theta}
\greeknewcommand{\Pib}{\overline\Pi}
\greeknewcommand{\Lab}{\bar\Lambda}
\greeknewcommand\Sib{\bar\Si}
\greeknewcommand\Xib{\bar\Xi}

\greeknewcommand{\da}{{\dot{\alpha}}}
\greeknewcommand{\db}{{\dot{\beta}}}
\greeknewcommand{\dd}{{\dot{\delta}}}
\greeknewcommand{\dep}{{\dot{\epsilon}}}
\greeknewcommand{\de}{{\dot\varepsilon}}
\greeknewcommand{\dg}{{\dot{\gamma}}}
\greeknewcommand{\dm}{{\dot{\mu}}}
\greeknewcommand{\dv}{\dot\varphi}

\greeknewcommand{\hcq}{{\hat{\cal Q}}}
\greeknewcommand{\hB}{\hat{B}}
\greeknewcommand{\hW}{\hat{W}}
\greeknewcommand{\psih}{{\hat{\psi}}}
\greeknewcommand{\psibh}{\hs{1mm}\hat{\hs{-1mm}\psib}{}}
\greeknewcommand{\omh}{\hat{\omega}}
\greeknewcommand{\hK}{\widehat{K}}
\greeknewcommand{\omt}{\tilde{\omega}}
\greeknewcommand{\Omt}{\tilde{\Omega}}
\greeknewcommand{\hh}{\tilde{h}}
\greeknewcommand{\what}{\widehat}
\greeknewcommand{\wti}{\widetilde}
\greeknewcommand{\wGm}{{\widehat{\Gamma}}}

\greeknewcommand\undal{{\underline{\alpha}}}
\greeknewcommand\undel{{\underline{\delta}}}
\greeknewcommand\undbt{{\underline{\beta}}}
\greeknewcommand\undgm{{\underline{\gamma}}}
\greeknewcommand\unddt{{\underline{\delta}}}
\greeknewcommand\undep{{\underline{\epsilon}}}
\greeknewcommand\undsi{{\underline{\sigma}}}
\greeknewcommand\undph{{\underline{\varphi}}}
\greeknewcommand\undSi{{\underline{\Sigma}}}
\greeknewcommand\undOm{{\underline{\Omega}}}

\@ifundefined{goth}{\newfont\goth{eufm10 scaled\magstep1}}{}
\nc{\A}{{\!A}}%			just a little closer...
\nc{\bA}{{\bar{A}}}
\nc{\bD}{{\bar{D}}}
\nc{\BF}{\xb F}
\nc{\bcd}{\xb\cd}
\nc{\bch}{\overline{\cal H}}
\nc{\bfA}{{\bf A}}
\nc{\bfD}{{\bf D}}
\nc{\bff}{{\bf f}}
\nc{\bfu}{{\bf u}}
\nc{\bfv}{{\bf v}}
\nc{\bfW}{{\bf W}}
\nc{\bh}{{\overline H}}
\nc{\bH}{\xb H}
\nc{\bi}{{\bar{\imath}}}
\nc{\bj}{{\bar{\jmath}}}
\nc{\bk}{{\bar{k}}}
\nc{\bK}{{\bar{K}}}
\nc{\bl}{{\bar{l}}}
\nc{\bp}{{\bar{p}}}
\nc{\bQ}{{\bar{Q}}}%			better use overline ??
\nc{\bq}{{\bar{q}}}
\nc{\br}{{\bar{r}}}
\nc{\bR}{{\xb R}}
\nc{\bs}{{\bar{s}}}
\nc\bS{\xb S}%				see below
\nc\bT{{\hskip.1ex\overline{\hskip-.1ex T}}}%
\nc{\bU}{{\bar U}}
\warnred{bv}{bar V}%	option french defines \bv (?)
\nc\bV{\xb V}
\nc\bW{\xb W}
\nc{\bw}{{\overline{W}}}
\nc{\bcw}{{\overline{\cw}}}
\nc{\bx}{{\overline{X}}}
\nc{\bX}{{\xb{X}}}
\nc{\by}{{\overline{Y}}}
\nc{\bz}{{\bar{z}}}
\nc{\fb}{{\bar f}}%		see also \barf
\nc{\ca}{{\cal A}}
\nc{\cb}{{\cal B}}
\warnred{cc}{\cal\space C}%	attention w/ letter.sty \cc
\nc{\cd}{{\cal D}}
\nc{\ce}{{\cal E}}
\nc{\cf}{{\cal F}}
\nc{\cg}{{\cal G}}
\nc{\ch}{{\cal H}}
\nc{\ci}{{\cal I}}
\nc{\cj}{{\cal J}}
\warnred{ck}{\cal\space K}%		\ck is (was) german "ck
\nc{\cl}{{\cal L}}
\nc{\cm}{{\cal M}}
\nc{\cn}{{\cal N}}
\nc{\co}{{\cal O}}
\nc{\cp}{{\cal P}}
\nc{\cq}{{\cal Q}}
\nc{\cR}{{\cal R}}%	\cr is carriage return. Anyway, \cX would be
\nc{\cs}{{\cal S}}%		more logical than \cx for all of those
\nc{\ct}{{\cal T}}
\nc{\cu}{{\cal U}}
\nc{\cv}{{\cal V}}
\nc{\cw}{{\cal W}}
\nc{\cx}{{\cal X}}
\nc{\cy}{{\cal Y}}
\nc{\cz}{{\cal Z}}
\nc{\fs}{{\bf f}}
\nc{\rd}{{\rm d}}
\nc{\re}{{\rm e}}
\nc{\rg}{{\rm g}}
\nc{\rA}{{\rm A}}
\nc{\rD}{{\rm D}}
\nc{\shalf}{\sm{1}{2}}%		OBSOLETE
\nc{\undA}{{\underline{A}}}
\nc{\undB}{{\underline{B}}}
\nc{\undC}{{\underline{C}}}
\nc{\prt}{{\partial}}
\nc\LRA{\Longleftrightarrow}%			or: \mbox{$\iff$}
\nc\impl{{\Rightarrow}}%				IMPLIES, THEREFORE
\nc\imag{\mathop{\Im\!m}}
\nc\Lie{\mathop{\rm Lie}}%			DIAG.MATRIX
\nc\eln{\mathop{\ell\kern-.1em n}}
\nc\real{\Re\!e\,}
\nc\msh{\mbox{$\frac12$}}%				MY SMALL HALF
\nc\mynparallel{\not{\!\|\,}}%			(w/ mssymb use nparallel)
\nc\sinc{{\rm sinc}}
\nc\tr{\mathop{\rm tr}}
\nc\Tr{\mathop{{\rm Tr}}}
\nc\Div{\partial\cdot}%					DIVERGENCE
\nc\dslash{\!{\not\!\partial}}
\nc\dsb{\!{\not\!\bar\partial}}%	BELIEVE ME OR NOT: THE {} IS NECESSARY!!
\nc\FT{{\cal F\!T}}
\nc\mybox{\square}%			{\fbox{\quad}} if no mssymb/AMS-TeX

\nc\gf{{\gamma^5}}
\nc\unity{1\hspace{-0.25em}{\rm l}}

\nc\crc[1]{{\begin{picture}(11,9)(0,1)\put(5.5,4.5){\circle{10}}\put(2,0.8){\elvsf #1}\end{picture}}}%

\nc\bwt[1]{{\overline{\wt{#1}}}}
\nc\lra[1]{\stackrel\leftrightarrow{#1}}%			<-> on s.th.
\nc\vect[1]{\stackrel\rightarrow{#1}}%				LONG VECTOR
\nc\vecto[1]{\stackrel\longrightarrow{#1}}%			LONGER VECTOR
\nc\wt{\widetilde}%						MY PREFERENCE
\nc\xb[1]{{\,\overline{\!#1}}}%				EXTENDED (LARGE) BAR
\nc{\sm}[2]{\hbox{\footnotesize$\displaystyle\frac{#1}{#2}$}}
\nc\sms[2]{\sm{#1}{#2}\,}%				with small space added
\nc{\thalf}{\tm12}
\nc{\tm}[2]{{\mbox{${#1\over#2}$}\,}}%	TEXTSIZE FRACTION
\nc\bigexp[1]{{\mbox{\large$\rm e^{#1}$}}}
\nc\der[2]{{\partial{#1}\over\partial{#2}}}%			PART. DERIVATIVE
\nc{\del}[2]{{{\delta}_{#1}}^{#2}}
\nc\diag[1]{\mathop{\rm diag}\lr(){#1}}%			DIAG.MATRIX
\nc\eqnrf[1]{\eqn{} \rf{#1}}
\nc\hstext[1]{\hs\mbox{#1}\hs}%		text in eqn, e.g. x=1 \hstext{and} y=0
\nc\idx[1]{\int\!{\rm d}^{#1}x\,}%			N-DIM. INTEGRAL OVER X
\nc\inv[1]{{#1}^{-1}}%						INVERSE
\nc\ip[1]{\iota_{#1}}%					INTERIOR PRODUCT
\nc\lr[3]{{\left#1 #3 \right#2}}%			\lr[]{...} => (...)
\nc\p[1]{{\left(#1\right)}}%				\p{...} => (...)
\nc\mycase[1]{\left\{\begin{array}{ccl}#1\end{array}\right.}
\nc\mymat[1]{\lr(){\begin{array}#1\end{array}}}%	MATRIX
\nc\mydbleqn[3]{\bea\begin{array}{rcl}#1%	TWO EQNARRAYS SIDE BY
 \end{array}#2\begin{array}{rcl}#3\end{array}\ena}%	SIDE (2.ARG EG. \label)
\nc\mea{\@ifnextchar[{\marray}{\marray[$\displaystyle\tabskip\z@{\@@@}$&
\global\@eqcnt\@ne\hskip2\arraycolsep\hfil${\@@@}$\hfil&
\global\@eqcnt\tw@\hskip2\arraycolsep$\displaystyle\tabskip\z@{\@@@}$]}}%
\def\marray[#1]#2{\stepcounter{equation}\let\@currentlabel=\theequation
\global\@eqnswtrue\global\@eqcnt\z@\tabskip\@centering
$$\let\@@@=##%	THIS SEEMS TO BE NECESSARY FOR DEFINITION OF \MEA
\let\\=\@eqncr\halign to \displaywidth\bgroup\@eqnsel\hskip\@centering
\tabskip\z@#1\hfil\tabskip\@centering&\llap{##}\tabskip\z@\cr
#2\@@eqncr\egroup\global\advance\c@equation\m@ne$$\global\@ignoretrue}%
\nc\set[1]{\lr\{\}{#1}}%				PUT ARG. IN BRACES
\nc\sslash{\,/\hspace{-1.5ex}}%			feynman slash
\nc\Slash{\,/\hspace{-1.3ex}}%			for small letters
\nc\SLASH[1]{\,/\hspace{-1.#1ex}}%		for any letters (\SLASH4w)
\nc\SU[1]{{\mbox{SU$(#1)$}}}
\nc\un[1]{{\underline{#1}}}
\nc\var[2]{{\delta{#1}\over\delta{#2}}}
\nc\BBB[1]{{\Bbb #1}}%					Bug in mssymb.tex
\warnred{I}{implement BBB replacement}
\def\I#1{{\rm I\!#1}}%					To make double bar
\nc\BB{\overline B}
\nc\BG{\overline G}
\nc\CC{\mbox{\rm C\hspace{-0.55em}\sf I~}}
\nc\DD{\I D}
\nc\HH{\I H}
\nc\KK{\I K}
\nc\NN{\I N}
\nc\QQ{{\sf l\hspace{-0.4em}\rm Q}}
\nc\RR{\I R}
\nc\WW{{\sf \backslash\!\!W}}	%{{\sf V\!\!V}}
\nc\ZZ{{\sf Z\!\!Z}}

\nc\eu[2]{\eps^{#1 #2}\,}%				LOWER INDICES
\nc\sig[3]{\sigma^{#1}_{#2#3}}%				SIGMA w/ INDICES

\nc\dlb{{\crc B}}%			BOSONIC DEGREES OF FREEDOM
\nc\dlf{{\crc F}}%			FERMIONIC DEGREES OF FREEDOM

\nc\0{|_{\th=0}}%	NB: \Big\0 gives big | but {...} on baseline :-(
\nc\ind[1]{_{\rm #1}}%				INDEX IN ROMAN
\nc\dk[1]{_{(#1)}}%				LOWER INDEX IN PARENTHESIS
\nc\uk[1]{^{(#1)}}%				UPPER INDEX IN PARENTHESIS
\nc\downup[2]{_{#1}^{{\phantom{#1}}#2}}%	LOWER,THEN UPPER INDEX
\nc\du[2]{{}_{#1}{\kern-.1em}^{#2}}
\nc\ud[2]{{^{#1}}{\kern-.1em}_{#2}}%
\nc\dud[3]{{}_{#1}\ud{#2}{#3}}
\nc\udu[3]{{}^{#1}\du{#2}{#3}}
\nc\up[1]{{}^{#1}\!}%		UPPER INDEX to be put closer to next symbol

\nc\zerobox[2]{{\raisebox{#1}[0pt][0pt]{$\scriptstyle #2$}}}%
\nc\zerozerobox[1]{\zerobox{0pt}{#1}}%
\nc{\lsym}[1]{{\mathop{#1}\limits_{\hbox{\large$\smile$}}}}
\nc{\sym}[1]{{\mathop{#1}\limits_{\scriptstyle\smile}}}
\nc\zsym[1]{ \zerozerobox{  \mathop{#1}\limits_{\zerobox{.5ex}\smile}  } }

\nc\lsy[1]{{%
\setbox\@tempboxa\hbox{$\scriptstyle #1$}
\@tempcnta\wd\@tempboxa
\divide\@tempcnta\unitlength
\@tempcntb\@tempcnta
\divide\@tempcntb by 2\relax
\advance\@tempcnta -4\relax
\begin{picture}(0,0)
\put(\@tempcntb,-1){\oval(\@tempcnta,5)[b]}
\end{picture}
\usebox\@tempboxa
}}

\nc\gp{\up g\Phi}
\nc\gA{ \,\up g\!A}

\nc\vi[1]{\cv\ind{#1}}%		potential (WITH INDEX)
\nc\li[1]{\cl\ind{#1}}%		lagrangien (WITH INDEX)
\nc\lcin{\li{cin}}%				cinetique
\nc\lmsq{\li{mini~SUSY~QED}}%			mini susy qed
\nc\lym{\li{YM}}%				Yang-Mills (cinetic)
\nc\lmat{\li{mat}}%				Matter
\nc\lmass{\li{mass}}%				mass
\nc\lpot{\li{pot}}%				Potential
\nc\lcu{\lcin^{\rm gauge}}%			cin.gauge [U(1)]
\nc\lcm{\lcin^{\rm mat}}%			cin. materie []
\nc\lcl{\lcin^{\rm lin}}%			cin. lineaire []
\nc\lpm{\lpot^{\rm mat}}%			pot. matiere [A\chi^2]
\nc\inth[1]{\int\!{\rm d}^#1\th\,}%		THETA INTEGRATION --- WARNING:
\nc\bpp{\lr(){\Phib\Phi}}
\nc\krs{\kappa\dk{r)(s}}%	kappa w/ (indices)
\nc\kp{K(\phi,\phib}%			KAHLER POTENTIAL \kp) or \kp;L)
\nc\HHb{\overline\HH}
\nc\Phid{\Phi^\dagger}
\nc\lieco[1]{\pounds_{#1}}	% covar lie abl.
\nc\UK{{$\rm U_K(1)$}}%					U(1)_K
\nc\warg{{\{\Phi_k\}}}	% args
\nc\stw{{\sin\theta_W}} % weinberg angle
\nc\ctw{{\cos\theta_W}} % weinberg angle
\remark{
}
\nc\CR{\nonumber\\}
\nc\nn{\nonumber}
\nc\numbertwo{\CR[-1.5ex]\\[-1.5ex]\nn}%		NUMBER IN BETWEEN 2 EQNS
\nc\lbl[1]{\label{eq:#1}}
\nc\rf[1]{(\ref{eq:#1})}
\def\section{\@ifstar{\subsection*}{\subsection*}}
\newcommand{\hD}{\hat \bfD}
\newcommand{\B}{\bar\D}
\newcommand{\D}{{\cal D}}                            % d
\newcommand{\w}{\omega}
\newcommand{\tD}{\tilde \D}
\newcommand{\eh}[1]{\rule{#1 cm}{0cm}}  %horizontal space
\newcommand{\drd}{\prt}
\newcommand{\BA}{\overline{A}}
\newcommand{\BM}{\overline{M}}
\newcommand{\BR}{R^\dagger}

\newcommand{\BX}{\overline{X}}
\newcommand{\BW}{\overline{W}}
\newcommand{\sih}[1]{\sigma^#1}
\newcommand{\sib}[1]{\sigma_#1}
\newcommand{\sibh}[1]{{\bar \sigma}^#1}
\newcommand{\sibb}[1]{{\bar \sigma}_#1}
\newcommand{\e}[2]{e\du{#1}{#2}}        %e_m^a
\newcommand{\E}[2]{E\du{#1}{#2}}        %E_m^a
\def\DC{\mbox{\large$\gm$}{\kern-.1ex}}
\begin{document}
%% include{t}%	title, abstract, toc
%% include{1}%	pref
\mynewpage
%
%	Title & abstract
%
%	for "U(1) supergravity" --- created 30-4-97
%
\def\CPT#1#2{#1}%	CHANGE TO #2 FOR SPECIAL CPT-PREPRINT-LAYOUT
\pagenumbering{roman}%	sets page no. = 1
\begin{titlepage}
\makeatletter
\title{\CPT{}{\vspace*{2cm}}\hbox to 0pt
{Higher--derivative supergravity in U(1) superspace}}
\author{R\'egis Le D\^u\thanks{allocataire M.\,E.\,S.\,R.; ~
E--mail: \tt ledu@cpt.univ-mrs.fr
}%
\CPT{\\
\small\em Centre de Physique Th\'eorique\thanks
{Unit\'e Propre de Recherche 7061},
C.N.R.S. Luminy, case 907, F--13288 Marseille}{}%
}
\date{}
\maketitle
\vfill
\begin{abstract}
The complete structure of curvature squared terms is analyzed in the context of chirally extended supergravity, with special emphasis on the gravitationally induced Fayet--Iliopoulos D--term. Coupling of (chiral) matter is discussed in relation with a possible extension to $U(1)$ supergravity of the equivalence mechanism between $\cR+\al\cR^2$ and General Relativity coupled to a scalar.
\end{abstract}
%\tableofcontents
\vfill
\vfill
\phantom{Keywords: }gauge theories, supersymmetry breaking.\\[4ex]
%\rightline
%\rightline
%{CPT--97/P.3500}\\[2ex]
%{version May 1997}\\[2ex]
%\rightline
%{hep--th/9705XXX}\\[3ex]
~\par
%anonymous ftp or gopher: cpt.univ-mrs.fr
\vfill
\end{titlepage}
\pagenumbering{arabic}
%% endinput
\mynewpage

%***************
%  INTRODUCTION
%***************

\section{Introduction}

Higher derivative supergravity theories \cite{FZ78}, \cite{KTN78}, \cite{CFGPP86}, \cite{The86}, \cite{RWZ90} have been proposed as messengers for supersymmetry breaking through gravitational effects \cite{CFG88}. More recently, the equivalence of $\cR+\cR^2$ theories to gravity coupled to a scalar \cite{Whi84}, in its supersymmetric version \cite{Cec87}, has been advocated to provide a supersymmetry breaking device as well \cite{HOW96f}, \cite{HOW96a}, \cite{HOW96e}.\\
All these scenarios are based on traditional supergravity. On the other hand, in the presence of a chiral abelian gauge structure, known as $U(1)$ supergravity \cite{How82}, \cite{Mul86} a supergravity induced D--term appears naturally \cite{Fre77}, \cite{WN78}.

In the present paper, we describe explicitly the complete structure of curvature squared terms in $U(1)$ supergravity, and discuss possibilities to extend the scheme of \cite{HOW96f} to this case.\\
In the first chapter we review shortly General Relativity \cite{Ste78} and Whitt's mechanism \cite{Whi84}. After an outline of $U(1)$ superspace and the construction of the pure $U(1)$ supergravity action in chapter 2 we turn to the complete description of curvature squared actions for $U(1)$ supergravity. Our description is based on methods of superspace geometry as reviewed in \cite{GG97}. The paper closes with a discussion of matter coupling to $U(1)$ supergravity with curvature squared terms.

%\mynewpage

%*********************
%  GENERAL RELATIVITY
%*********************

\section{General Relativity with curvature squared terms}

It is known that pure General Relativity is a nonrenormalizable theory \cite{HV74}\cite{DN74}. Adding quadratic terms in the curvature tensor 
allows to construct renormalizable actions \cite{Ste77}. The most general action which contains fourth order derivatives can be written as
\be
\cs = \gamma \int d^4x \sqrt{-g}
      \lr(){\cR+\al'\cR^2+\beta'\cR^{mn}\cR_{mn}+c\cR^{mnpq}\cR_{mnpq}}~,
\label{z1}
\ee
where $\gamma$ is related to the gravitational constant, resp. the Planck mass,
\be
\gamma = {-1\over 16\pi G_N} = -\frac 12 M_P^2~,
\label{z2}
\ee
whereas $\gamma\alpha'$, $\gamma\beta'$ and $\gamma c$ are dimensionless. Taking into account the Gauss--Bonnet combination
\be
\cs_{GB} =  \int d^4x \sqrt{-g} \lr(){\cR^2-4\cR^{mn}\cR_{mn}+\cR^{mnpq}\cR_{mnpq}}~,
\label{z3}
\ee
which, as a topological invariant, does not depend on the metric and, as a consequence, does not contribute to the equations of motion, this action may be written as
\be
\cs'=\cs-c\gamma\cs_{GB} = \gamma \int d^4x \sqrt{-g}
                     \lr(){\cR+\al\cR^2+\beta\cR^{mn}\cR_{mn}}~,
\label{z4}
\ee
with constants $\al=\al'-c$, $\beta=\beta'+4c$. This theory describes \cite{Ste78} \cite{BC83} the graviton together with a massive spin two "poltergeist" and a massive physical scalar field. For $\beta=0$ the poltergeist decouples \cite{Cec87} and the action 
\be
\cs_{\cR+\cR^2} = \gamma \int d^4 x \sqrt{-g} \lr(){\cR+\al\cR^2}~,
\label{z5}
\ee
which, however, is no longer renormalizable \cite{Ste77}, describes a graviton coupled to a massive scalar field \cite{Whi84}. Following \cite{Whi84}, one starts from the action
\be
\cs_{\cR+\phi} = \gamma \int d^4 x \sqrt{-g} \lr(){\cR+2\al\phi\cR-\al\phi^2}~.
\label{z6}
\ee
On the one hand, varying with respect to $\phi$ reproduces the action (\ref{z5}). On the other hand, performing a Weyl rescaling
\be
g_{mn} = \lr(){1+2\al\phi}^{-1} g'_{mn}~,
\label{z7}
\ee
yields
\be
\cs_{\cR+\phi} = \gamma \int d^4 x \sqrt{-g'}  
                  \lr(){\cR'
                       +6\al^2 (1+2\al\phi)^{-2} \drd'^m\phi\drd'_m\phi
                       -\al (1+2\al\phi)^{-2} \phi^2
                       }~,
\label{z8}
\ee
where the presence of the massive scalar field is manifest. It is in this sense that curvature squared gravity is said to be equivalent to General Relativity coupled to a scalar field \cite{Whi84}. 
%\mynewpage
%
%*****************
% U(1) SUPERSPACE
%*****************

\section{U(1) superspace and supergravity}

Supergravity is described in terms of the vielbein field $\e ma(x)$ and the Rarita--Schwinger field $\psi\du{m}{\al} (x)$ together with a set of auxiliary fields%
\footnote{This set of auxiliary fields corresponds to the so called "old--minimal" formulation. Other choices are possible, as for instance  "new--minimal" and "non--minimal" formulations, but they will not be considered here.}
$M, \BM$ and $b_a$, i.e a complex scalar and a vector. The latter are necessary to close the algebra of local supersymmetry transformations. In conventional supergravity which is given as the supersymmetric generalization of the curvature scalar they describe non propagating degrees of freedom.\\
On the other hand, in supersymmetric versions of theories with curvature squared terms those fields acquire derivatives and become propagating fields as well.\\
In the present paper we will extend this scenario to the case of $U(1)$ supergravity, describing an enlarged multiplet with additional components $A_m$, $\lambda^\al$ and $\bf D$. This theory has an additional gauged chiral $U(1)$ symmetry with $A_m$ as gauge potential, $\lambda^\al$ the gaugino superpartner and $\bf D$ the auxiliary field. One of its interesting features is that it allows the construction of a particular supersymmetric Fayet--Iliopoulos \cite{FI74} term in the context of supergravity \cite{WN78}.\\
The superspace formulation of $U(1)$ supergravity \cite{How82} \cite{Mul86} is a generalization of that of conventional supergravity \cite{WeB83}~: in addition to the Lorentz group in the superspace structure group it contains a chiral $U(1)$. In order to be more explicit we review shortly the salient features of $U(1)$ superspace geometry, following closely refs. \cite{BGG90}, \cite{GG97}.
The basic objects are the vielbein $E^A$, the Lorentz connection $\phi\du BA$ and the $U(1)$ connection $A$. They are all one--forms in superspace~:
\bea
E^A &=& dz^M \E MA~,
\label{z9} \\
\phi\du BA &=& dz^M \phi\du{MB}{A}~,
\label{z10} \\
A &=& dz^M A_M~.
\label{z11}
\ena
Correspondingly, one defines the torsion $T^A$, the Lorentz curvature $R\du BA$, and the $U(1)$ fieldstrength $F$~:
\bea
T^A &=& dE^A + E^B\phi\du BA + \w(E^A) E^A A~,
\label{z12} \\
R\du BA &=& d\phi\du BA + \phi\du BC \phi\du CA~,
\label{z13} \\
F &=& dA~,
\label{z14}
\ena
which are two--forms in superspace. The chiral weights of the vielbein are defined to be 
\be
\w(E^a)=0~,~\w(E^\al)=1~,~\w(E_\da)=-1~.
\label{z15}
\ee
The basic covariant superfields which completely describe torsion, curvature and $U(1)$ fieldstrength are
\be
R~, \eh{.3} 
\BR~, \eh{.3}
G_a~, \eh{.3} 
W_{\lsym{\al\beta\gamma}}~, \eh{.3} 
\BW^{\lsym{\da\db\dg}}~, \eh{.3}
\mbox{and} \eh{.3}
X^\al~, \eh{.3}
\BX_\da~.
\label{z16}
\ee
Component fields are defined as lowest components of superfields in the usual way
\be
\E ma|=\e ma(x)~, \eh{.3}
\E m\al|=\frac{1}{2}\psi\du{m}{\al}(x)~, \eh{.3}
A_m|=A_m(x)~,
\label{z17}
\ee
for the gauge fields. In particular the metric tensor is defined as $g_{mn}=\e ma\e nb \eta_{ab}$, with $\eta_{ab}=\mbox{diag}\lr(){-,+,+,+}$. Moreover one has the usual definitions
\be
R|=-\frac{M}{6}~, \eh{.3}
\BR|=-\frac{\BM}{6}~, \eh{.3}
G_a|=-\frac{b_a}{3}~,
\label{z18}
\ee
in the gravity sector and
\be
X_\al| = -i \lambda_\al~, \eh{0.3}
\BX^\da| = i \bar\lambda^\da~, \eh{0.3} 
\cd^\al X_\al|= -2 \bfD~,
\label{z19}
\ee
in the $U(1)$ sector.
We also define a gauge potential% 
\footnote{This definition takes into account the constraint~:~$F\ud\db\al=-3G\ud\db\al$ which is used in \cite{BGG90}.}
 ${\tilde A}_m$ which is related to $A_m$ by 
\be
{\tilde A}_m \equiv A_m +\frac {i}{2} b_m~,
\label{z20}
\ee
and which will be used as the basic component field from now on. Correspondingly, we define the $U(1)$ covariant derivative $\tD_m X$
\be
\tD_m X = \drd_m X+\w(X){\tilde A}_m X = \D_m X +\frac {i}{2} w(X) b_m X~.
\label{z21}
\ee
The supergravity action is given in compact form as the volume element of $U(1)$ superspace, i.e%
\footnote{Action and lagrangian are related by the relation~:~$\cs=\int d^4x \cl$.}%
\be
\cs_1 = -3 \int E~.
\label{z22}
\ee
The corresponding component field expression is most conveniently extracted from the generic lagrangian \cite{BGG90}~:
\be
e^{-1} \cl_1 = -\frac{1}{4} \D^2 r|
               - r|(\BM+\bar\psi_m\bar\sigma^{mn}\bar\psi_n)  
               + \frac{i}{2}  \lr(){\bar\psi_{m\da}\bar\sigma^{m\da\al}}\D_\al r|  
               + hc~,
\label{z23}
\ee
with the choice
\be
r = -3 R~.
\label{z24}
\ee
Using the explicit component field form of $\D_\al R|$, $\D^2 R|$, one obtains
\bea
e^{-1}\cl_1 &=&  -\frac{M_P^2}{2} \cR 
                 +\frac{1}{2}  \epsilon^{mnpq} \lr(){\bar\psi_m\sibb n \tD_p\psi_q 
                                                - \psi_m\sib n \tD_p \bar\psi_q  }
                 -\frac{M_P^2}{3}\lr(){M\BM - b^a b_a} \CR
            & &
                 +\frac 12  \lr(){ \bar\psi_m\sibh m \lambda 
                                  - \psi_m\sih m \bar\lambda }M_P 
                 +\bfD M_P^2~.
\label{z25}
\ena
Clearly, this lagrangian exhibits the usual Einstein term together with a kinetic term for the gravitino and the auxiliary field terms for $M$, $\BM$ and $b_a$ as in usual supergravity. One of the new features due to the chiral $U(1)$ is the appearance of ${\tilde A}_m$ in the covariant derivative of the gravitino field
\be 
\tD_m\psi\du n\al = \drd_m\psi\du n\al + \psi\du n\beta\w\du{m\beta}{\al} + \psi\du n\al {\tilde A}_m~.
\label{z26}
\ee
Moreover there is a term linear in $\bfD$, which clearly shows that this theory in itself cannot be complete. It is completed in adding a kinetic term for the $U(1)$ gauge multiplet. In this case the term linear in $\bfD$ can play the role of a Fayet--Iliopoulos term \cite{WN78}.
The superfield action is defined as 
\be
\cs_{X^2} =     \frac 18 \int {E \over R} X^\al X_\al 
            ~+~ \frac 18 \int {E \over \BR} \BX_\da \BX^\da~.
\label{z27}
\ee
Taking $r=\frac{1}{4} X^\al X_\al$ in the generic construction (\ref{z23}), one derives the component field expression
\bea
e^{-1}\cl_{X^2} = \frac 12 \bfD^2 + {\tilde F}^{mn} {\tilde F}_{mn}
                  -\frac{i}{2} \lambda\sigma^m\tD_m\bar\lambda
                  -\frac{i}{2} \bar\lambda\bar\sigma^m\tD_m\lambda
                  +\mbox{other fermionic terms}~.
\label{z28}
\ena
These two lagrangians are separately invariant under the following supersymmetry transformations (as derived from superspace geometry in the usual way)
\bea
\delta_\xi \e ma &=& i (\xi\sih a \bar\psi_m + \bar\xi\sibh a \psi_m)M_P^{-1}~, 
\label{z29} \\
\delta_\xi \psi\du m\al &=& \lr(){2\tD_m\xi^\al
                            -i\xi^\al b_m 
                            -\frac i3 (\xi\sih a\sibb m)^\al b_a 
                            +\frac i3 (\bar\xi\sibb m)^\al M} M_P~,
\label{z30} \\
\delta_\xi \bar\psi_{m\da} &=& \lr(){2\tD_m\bar\xi_\da
                               +i\bar\xi_\da b_m
                               +\frac i3 (\bar\xi\sibh a\sib m)_\da b_a
                               +\frac i3(\xi\sib m)_\da\BM} M_P~,
\label{z31} \\
\delta_\xi M &=&  -2i(\xi\lambda)
                 \lr(){  +4(\xi\sigma^{mn}\tD_m\psi_n)
                         +i(\psi_m\xi)b^m
                         +i(\bar\psi_m\sibh m\xi)M} M_P^{-1}~,
\label{z32} \\
\delta_\xi \BM &=& 2i(\bar\xi\bar\lambda)
                   \lr(){ +4(\bar\xi{\bar\sigma}^{mn}\tD_m\bar\psi_n)
                          -i(\bar\psi_m\bar\xi)b^m
                          -i(\bar\xi\sibh m\psi_m)\BM} M_P^{-1}~,
\label{z33} \\
\delta_\xi b_a &=&  \lr(){ \frac 12 (\xi\sib a{\bar\sigma}^{mn}\tD_m\bar\psi_n)
                          +\frac 12 (\xi\sib a{\bar\sigma}^{mn}\bar\psi_n)b_m } M_P^{-1} \CR
               & &  \lr(){-\frac 32 (\xi\sigma^{mn}\sib a\tD_m\bar\psi_n)
                          -\frac 32 (\xi\sigma^{mn}\bar\psi_n)b_m             } M_P^{-1} \CR
               & &  \lr(){-\frac i2 \e am(\xi\sih d\bar\psi_m)b_d
                          +\frac i2 \e am(\xi\psi_m)\BM                       } M_P^{-1} \CR
               & & +i(\xi\sib a\bar\lambda)+ h.c~,
\label{z34} \\
\delta_\xi \bfD &=& (\bar\xi\sibh m\tD_m\lambda)
                   -(\xi\sih m\tD_m\bar\lambda)
                   -\frac i2 (\xi\sih m\bar\lambda + \bar\xi\sibh m\lambda)b_m \CR            
                & &+\frac 12 (\psi_m\sigma^{kl} \sih m\bar\xi-\bar\psi_m {\bar\sigma}^{kl} \sibh m\xi)
                             (2i{\tilde F}_{kl}M_P^{-1}+i\psi_k\sib l\bar\lambda M_P^{-2} +i\bar\psi_k\sibb l\lambda M_P^{-2}) \CR 
                & &+\frac i2 (\bar\psi_m\sibh m\xi+\psi_m\sih m\bar\xi)\bfD M_P^{-1}~,
\label{z35} \\
\delta_\xi {\tilde A}_m &=& \frac{1}{2} (\bar\lambda\sib m\xi + \lambda\sibb m\bar\xi)~,
\label{z36} \\
\delta_\xi \lambda^\al &=& (\xi\sigma^{nm})^\al
                           (2i{\tilde F}_{nm}
                            +\psi_n\sib m\bar\lambda M_P^{-1} 
                            +\bar\psi_n\sibb m\lambda M_P^{-1}) 
                           +i\xi^\al\bfD~,
\label{z37} \\
\delta_\xi \bar\lambda_\da &=& (\bar\xi{\bar\sigma}^{nm})_\da
                               (2i{\tilde F}_{nm}
                                +\psi_n\sib m\bar\lambda M_P^{-1}
                                +\bar\psi_n\sibb m\lambda M_P^{-1})
                               -i\bar\xi_\da\bfD~.
\label{z38}
\ena
Here the $U(1)$ fieldstrength
\be
{\tilde F}_{kl} = \drd_k {\tilde A}_l -\drd_l {\tilde A}_k~,
\label{z39}
\ee
and the covariant derivatives
\bea
\tD_m\lambda^\al &=& \drd_m\lambda^\al 
                    + \lambda^\beta\w\du{m\beta}{\al} 
                    + \lambda^\al {\tilde A}_m~,
\label{z40}\\
\tD_m\bar\lambda_\da &=& \drd_m\bar\lambda_\da 
                   + \bar\lambda_\db \w\dud{m}{\db}{\da} 
                   - \bar\lambda_\da {\tilde A}_m~. 
\label{z41}
\ena
as well as
\bea
\tD_m\xi^\al &=& \drd_m\xi^\al + \xi^\beta\w\du{m\beta}{\al} + \xi^\al {\tilde A}_m~,
\label{z42}\\
\tD_m\bar\xi_\da &=& \drd_m\bar\xi_\da 
               + \bar\xi_\db \w\dud{m}{\db}{\da} 
               - \bar\xi_\da {\tilde A}_m~, 
\label{z43}
\ena
occur.\\
Observe that the sum of $\cl_1$ and $\cl_{X^2}$, which might be referred to as pure $U(1)$ supergravity, gives rise to a cosmological constant after diagonalization in the field $\bfD$, as discussed in \cite{WN78} \cite{Fre77} \cite{Zum75}. In our case this action provides the starting point for the discussion of curvature squared terms and diagonalization should only be performed afterwards.
%\mynewpage
%
%*************************************
%  HIGHER--DERIVATIVE IN SUPERGRAVITY
%*************************************

\section{Curvature squared terms and $U(1)$ supergravity}

As is well known \cite{FZ78} curvature squared terms in traditional supergravity are identified in the highest superfield components of the combinations $W^{\al\beta\gamma} W_{\al\beta\gamma}$, $G^a G_a$ and $R\BR$ of basic supergravity superfields. In the case of U(1) supergravity a number of modifications arise due to the presence of the $U(1)$ sector in the geometry, as explained in detail in \cite{GG97}. In order to fix our notations we shall consider here the three superspace actions%
\footnote{The combination $G^aG_a+2R\BR$ which appears in the second action is particularly convenient for discussion of the supersymmetric Gauss--Bonnet invariant.}
\bea
\cs_{W^2} &=& \int {E \over 2R} 
              \lr(){ W^{\lsym{\al\beta\gamma}} W_{\lsym{\al\beta\gamma}} } + h.c~,
\label{z44}\\
\cs_{G^2+2R\BR} &=& \int E \lr(){G^aG_a + 2R\BR}~,
\label{z45}\\
\cs_{R\BR} &=& \int -3 \lr(){36 R\BR}~.
\label{z46}
\ena 
Complete component field expressions can be evaluated in using the generic component field action (\ref{z23}) with the identifications, respectively,
\bea
r_{W^2} &=& W^{\lsym{\al\beta\gamma}} W_{\lsym{\al\beta\gamma}}~,
\label{z47}\\
r_{G^2+2R\BR} &=& -\frac{1}{8} \lr(){\B^2-8R} \lr(){G^aG_a + 2R\BR}~,
\label{z48}\\
r_{R\BR} &=& \frac{3}{8} \lr(){\B^2-8R} \lr(){36 R\BR}~,
\label{z49}
\ena
for the generic chiral superfield $r$.\\
In what follows we shall only discuss the purely bosonic contributions of these actions. Following \cite{GG97} one obtains
\bea
e^{-1} \cl_{W^2} &=&  \frac 18 \cw^{dc,ba} \cw_{dc,ba} 
                     +\frac 13 \lr(){F^{mn} F_{mn}}~,
\label{z50}\\
e^{-1} \cl_{G^2+2R\BR} &=&  -\frac{1}{8} \tilde\cR^{ba}\tilde\cR_{ba} 
                            +\frac{1}{96} \cR^2 
                            -\frac{1}{6} \bfD^2 
                            -\frac{1}{6} \lr(){ F^{mn}F_{mn} 
                                         + 2 {\tilde F}^{mn} {\tilde F}_{mn}}~,
\label{z51}\\
e^{-1}\cl_{R\BR} &=&   -\frac{3}{4} \lr(){\cR-2\bfD}^2
                       +\lr(){b^mb_m+\frac{1}{2} M\BM} \cR
                       -2\lr(){b^mb_m+2M\BM}           \bfD
\CR 
           & &   +3 \tD^m M\tD_m\BM 
                 -3 \lr(){\e am \tD_mb^a}^2
                 +i b^m\lr(){\BM\tD_m M-M\tD_m \BM} 
\CR
          & &    -\frac{1}{3} \lr(){(M\BM)^2+M\BM b^ab_a+(b^ab_a)^2}~,
\label{z52}
\ena
with the conventions
\bea
F_{mn} &=& {\tilde F}_{mn} + \frac{i}{2} B_{mn}~, 
\label{z53}\\
B_{mn} &=& \drd_m b_n-\drd_n b_m~,
\label{z54}
\ena
and where the Weyl tensor $\cw_{dc,ba}$, the Ricci tensor ${\tilde\cR}_{ca}$ and the curvature scalar $\cR$ are identified as usual in the decomposition of the Riemann tensor $\cR_{dc,ba}$~:
\be
\cR_{dc,ba} = \cw_{dc,ba}
           +\frac 12 \lr(){\eta_{db} \tilde\cR_{ca}
                          -\eta_{da} \tilde\cR_{cb}
                          -\eta_{cb} \tilde\cR_{da}
                          +\eta_{ca} \tilde\cR_{db} }
           +\frac{1}{12} \lr(){\eta_{db}\eta_{ca}-\eta_{da}\eta_{cb}}\cR~.
\label{z55}
\ee
With the first two lagrangians (\ref{z50}) and (\ref{z51}), one can obtain the supersymmetric version of the pure Gauss--Bonnet invariant (\ref{z3}) plus terms involving $\bfD^2$ and ${\tilde F}^{mn}{\tilde F}_{mn}$. These new contributions are due to the additional $U(1)$ sector, they can be cancelled by adding $\cl_{X^2}$. As a result, the pure super--Gauss--Bonnet combination is~:
\bea
\cl_{GB}   &=& 8\cl_{W^2}  + 16\cl_{G^2+2R\BR} + \frac{16}{3} \cl_{X^2}~.
\label{z56}
\ena
In terms of component fields this reproduces exactly the combination of equation (\ref{z3}), i.e~:
\bea
e^{-1} \cl_{GB} &=& \cw^{dc,ba} \cw_{dc,ba} 
                    -2 {\tilde\cR}^{ba}{\tilde\cR}_{ba} 
                    +\frac{1}{6} \cR^2~.
\label{z57}
\ena
In order to discuss the most general form of $U(1)$ supergravity with curvature squared terms we shall consider the combination%
\footnote{$a_i$ are real constants.}
\be
\cl_{tot}= a_1 \cl_1
          +a_2 \cl_{X^2}
          +a_3 \cl_{R\BR}
          +a_G \cl_{G^2+2R\BR}
          +a_W \cl_{W^2}
\label{z58}
\ee
with $\cl_1$ and $\cl_{X^2}$ defined in the previous section. Inspection of the individual contributions shows that diagonalization in the auxiliary field $\bfD$ of the $U(1)$ sector will introduce additional curvature scalar squared terms. More precisely, defining
\be
\hD \equiv \bfD+\frac{c}{2}\lr(){ a_1 + 3{a_3}\cR-4{a_3}M\BM-2{a_3}b^mb_m}~,
\label{z59}
\ee
with
\be
c = \lr(){-3a_3+\frac{a_2}{2}-\frac{a_G}{6}}^{-1}~,
\label{z60}
\ee
gives rises to the component field lagrangian 
\bea
e^{-1}
\cl_{tot} &=&    -\frac{a_1}{2} \lr(){1+{3a_3 c}} \cR
                 -\frac{a_1}{3} \lr(){1-{6a_3 c}} M\BM
                 +\frac{a_1}{3} \lr(){1+{3a_3 c}} b^ab_a 
\CR
          & &    +\frac{a_W}{8} \cw^{dc,ba} \cw_{dc,ba} 
                 -\frac{a_G}{8} {\tilde\cR}^{ba} {\tilde\cR}_{ba} 
                 +\lr(){-\frac{3a_3}{4}\lr(){1+{3a_3 c}}+\frac{a_G}{96}} \cR^2
\CR
          & &    +\lr(){a_2-\frac{a_G}{3}}  {\tilde F}^{mn}{\tilde F}_{mn}
                 +\frac{1}{6} \lr(){2a_W-a_G} F^{mn} F_{mn}
                 +3a_3           \tD^m M\tD_m\BM 
                 +c^{-1}         {\hat{\bfD}}^2 
\CR
          & &    -3a_3 \lr(){\e am \tD_mb^a}^2
                 +ia_3 b^m\lr(){\BM\tD_m M-M\tD_m \BM} 
\CR
          & &    +\frac{a_3}{2} \lr(){1+{12a_3 c}}  \cR M\BM
                 +a_3           \lr(){1+{3a_3  c}}  \cR b^ab_a
                 -\frac{a_1^2c}{4} 
\CR
          & &    -\frac{a_3}{3} \lr(){1+{12a_3 c}} \lr(){(M\BM)^2+M\BM b^ab_a}  
                 -\frac{a_3}{3} \lr(){1+{3a_3  c}} (b^ab_a)^2~.
\label{z61}
\ena
The Lorentz and $U(1)$ covariant derivatives appearing here are defined as
\bea
\tD_m M &=& \drd_m M + 2{\tilde A}_mM~,
\label{z62}\\
\tD_m b^a &=& \drd_m b^a + b^c \w\du{mc}{a}~,
\label{z63}
\ena
according to the chiral weights~:
\be
\w(M) = 2~, \eh{1} \w(b^a) = 0~.
\label{z64}
\ee
This is the (bosonic part) of the component field lagrangian which is relevant for the discussion of curvature squared terms in $U(1)$ supergravity. As a first observation consider the special case
\be
a_G = 2a_W~,
\label{z65}
\ee
which adjusts the relative factor between the squares of the Weyl and the Ricci tensors to that occurring in the Gauss--Bonnet combination. In this case the general action is specified to
\bea
e^{-1}
\cl_{tot} &=&    -\frac{a_1}{2} \lr(){1+{3a_3 c}} \cR
                 -\frac{a_1}{3} \lr(){1-{6a_3 c}} M\BM
                 +\frac{a_1}{3} \lr(){1+{3a_3 c}} b^ab_a 
\CR       & &    +\frac{a_W}{8} \lr(){ 
                               \cw^{dc,ba} \cw_{dc,ba} 
                              -2 {\tilde\cR}^{ba} {\tilde\cR}_{ba} 
                              +\frac{1}{6} \cR^2}
\CR
          & &    -\frac{3a_3}{4}\lr(){1+{3a_3 c}} \cR^2
                 +2c^{-1}\lr(){1+3a_3c}  {\tilde F}^{mn}{\tilde F}_{mn}
                 +3a_3           \tD^m M\tD_m\BM 
\CR
          & &    +c^{-1}         {\hat{\bfD}}^2 
                 -3a_3 \lr(){\e am \tD_mb^a}^2
                 +ia_3 b^m\lr(){\BM\tD_m M-M\tD_m \BM} 
\CR
          & &    -\frac{a_3}{3} \lr(){1+{12a_3 c}} \lr(){(M\BM)^2+M\BM b^ab_a}  
                 -\frac{a_3}{3} \lr(){1+{3a_3  c}} (b^ab_a)^2
\CR
          & &    +\frac{a_3}{2} \lr(){1+{12a_3 c}}  \cR M\BM
                 +a_3           \lr(){1+{3a_3  c}}  \cR b^ab_a
                 -\frac{a_1^2c}{4}~.
\label{z66}
\ena
This action is the $U(1)$ supergravity analogue of the case $\beta=0$ discussed for the non--supersymmetric case (\ref{z5}).\\
As an aside, note that in order to obtain the Gauss--Bonnet combination of curvature squared terms one has to cancel the additional $\cR^2$ term. This can be done by choosing either $a_3 = 0$ or $\lr(){1+3a_3 c}=0$. In the first case, the lagrangian is reduced to
\bea
e^{-1}
\cl_{tot} &=&  -\frac{a_1}{2}  \cR
               -\frac{a_1}{3}  M\BM
               +\frac{a_1}{3}  b^ab_a 
               +2c^{-1}        {\tilde F}^{mn} {\tilde F}_{mn}
               +c^{-1}         {\hat\bfD}^2
               -\frac{a_1^2c}{4}
\CR
         & &    +\frac{a_W}{8} \lr(){ 
                        \cw^{dc,ba} \cw_{dc,ba} 
                        -2 {\tilde\cR}^{ba} {\tilde\cR}_{ba} 
                        +\frac{1}{6} \cR^2}~,
\label{z67}
\ena 
and in the second case one obtains
\bea
e^{-1}
\cl_{tot} &=&   
                 -a_1 M\BM
                 -\frac{3a_3}{2} M\BM \cR
                 +3a_3           \tD^m M\tD_m\BM 
\CR
          & &    -3a_3        {\hat{\bfD}}^2 
                 -3a_3 \lr(){\e am \tD_mb^a}^2
                 +ia_3 b^m\lr(){\BM\tD_m M-M\tD_m \BM} 
\CR
          & &    +\frac{a_1^2}{12a_3} 
                 +a_3  \lr(){(M\BM)^2+M\BM b^ab_a}
\CR
& &    +\frac{a_W}{8} \lr(){ 
                        \cw^{dc,ba} \cw_{dc,ba} 
                        -2 {\tilde\cR}^{ba} {\tilde\cR}_{ba} 
                        +\frac{1}{6} \cR^2}~.
\label{z68}
\ena
Clearly, the first case describes a generalization of a supergravity action with a correctly normalized (for $a_1=1$) curvature scalar term.\\
The interpretation of the second case is more subtle in that a Weyl rescaling in $M\BM$ should be performed \cite{CFG88} \cite{HOW96f} to arrive at a correctly normalized Einstein term.\\
Finally, the coupling of curvature squared term to traditional supergravity can be recovered from equation (\ref{z61}) in simply switching off the $U(1)$ sector, i.e taking $a_2=0$, ${\tilde A}_m=0$ and substituting $\hD \equiv \frac{c}{2}\lr(){ a_1 + 3{a_3}\cR-4{a_3}M\BM-2{a_3}b^mb_m}$, which eliminates the c--dependence in (\ref{z61}).

%
%\mynewpage
%
%********************
% COUPLING TO MATTER
%********************

\section{Coupling to matter}
As chirally extended $U(1)$ supergravity provides a natural framework for a gravity coupled Fayet--Iliopoulos term \cite{Fre77}, \cite{WN78}, it is interesting to investigate couplings to chiral matter in this context. This discussion serves at the same time as a prerequisite for the generalization of the Whitt mechanism, as alluded to in the first section, to the case of $U(1)$ supergravity.

To begin with, we consider a single chiral superfield $\Phi$ of $U(1)$ weight $\w(\Phi)=\w$, and, correspondingly $\bar\Phi$ of $U(1)$ weight $\w(\bar\Phi)=-\w$. Evaluating the supersymmetric action
\be
\cs= a_1\cs_1
    +a_2\cs_{X^2}
    +a_4\int E f(\Phi,\bar\Phi) 
    +a_5\lr(){ \int {E\over 2R} \Phi^x
              +\int {E\over 2\BR} \bar\Phi^x}~,
\label{z69}
\ee
where $x$ is given in term of the chiral weight~:~$x={\w(R)\over\w(\Phi)}$ for $\w(\Phi)\ne 0$.
In terms of component fields, one finds, for the purely bosonic contribution,
\bea
e^{-1} \cl &=& \lr(){a_1+a_4 f(A,\BA)} 
               \lr(){-\frac{1}{2} \cR
                     -\frac{1}{3} M\BM
                     +\frac{1}{3} b^mb_m} 
\CR
            & & +\lr(){a_1+a_4 f(A,\BA) -\frac{3}{4} a_4 \w \lr(){f_A A+f_{\BA}\BA} }\bfD
\CR
            & & +\frac{a_2}{2} \bfD^2 
                + a_2 {\tilde F}^{mn} {\tilde F}_{mn}
\CR 
            & & +ia_4 b^m\lr(){f_A\tD_m A-f_{\BA} \tD_m\BA}
                +3a_4f_{A\BA} \lr(){ \tD^m A\tD_m\BA -F\BF}
\CR
            & & +a_4 \lr(){f_A MF + f_{\BA}\BM\BF}
                +a_5x\lr(){FA^{(x-1)} + \BF\BA^{(x-1)}}~,
\label{z70}
\ena
with the definitions
\be
\Phi|=A~, \eh{1} \D^2\Phi|=-4F~.
\label{z71}
\ee
Clearly, this provides a generalization of the Fayet--Iliopoulos term to the case of matter--coupled chirally extended supergravity, and possible applications to symmetry breaking mechanisms deserve further study.

On the other hand, this action is the starting point for the generalization of the Whitt's mechanism as well. In order to establish the relation between curvature squared $U(1)$ supergravity with its matter coupled counterpart, linear in the curvature scalar, we shall start from equation (\ref{z66}) with the particular choice $a_G=a_W=0$, i.e
\be
\cl=a_1\cl_1+a_2\cl_{X^2}+a_3\cl_{R\BR}~,
\label{z72}
\ee
which is a supersymmetric version of (\ref{z5}). As to the supersymmetric analogue of (\ref{z6}) we consider
\be
\cs=a_1\cs_1
   +a_2\cs_{X^2}
   +a_3\lr(){-3\int E \lr(){\Phi\bar\Phi+\Lambda+\bar\Lambda}
             +\int {E\over 2R}   \Lambda\Phi
             +\int {E\over 2\BR} \bar\Lambda\bar\Phi }~,
\label{z73}
\ee
which has the same appearance as the corresponding action in traditional supergravity \cite{Cec87}. However, in the present context a number of new features appear. In particular, for reasons of consistency with the $U(1)$ gauge structure, the $U(1)$ weights of the chiral superfields $\Phi$ and $\Lambda$ are determined to be~:
\be
\w(\Phi)=2~, \eh{1} \w(\Lambda)=0~.
\label{z74}
\ee
Evaluation of the purely bosonic part of this action in terms of component fields gives
\bea
e^{-1} \cl &=& \lr(){a_1+a_3\lr(){A\BA+B+\BB}} 
               \lr(){-\frac{1}{2} \cR
                     -\frac{1}{3} M\BM
                     +\frac{1}{3} b^mb_m} 
\CR
            & & +\lr(){a_1+a_3\lr(){1-\frac32 \w(\Phi)}A\BA
                                     + a_3\lr(){B+\BB}}\bfD
\CR
            & & +\frac{a_2}{2} \bfD^2 
                + a_2 {\tilde F}^{mn} {\tilde F}_{mn}
\CR 
            & & +ia_3 b^m\lr(){\BA\tD_m A-A\tD_m\BA+\tD_m B-\tD_m\BB}
                +3a_3\tD^m A\tD_m\BA
\CR
            & & +a_3\lr(){-3F\BF+\BA MF+A\BM\BF+FB+\BF\BB}
\CR
            & & +a_3\lr(){G(M+A)+\BG (\BM+\BA)-AB\BM-\BA\BB M}~,
\label{z75}
\ena
with the definitions~:
\bea
\Lambda| &=& B~, \eh{1}
\D^2 \Lambda| =-4G~.
\label{z76}
\ena
Taking into account the equations of motion%
\footnote{In superfield language this corresponds to varying (\ref{z73}) with respect to (the prepotential of) the chiral superfield $\Lambda$.}
\bea
A&=&-M~,
\label{z77}
\\
F&=&A\BM
     +\frac{1}{2}\cR
     -\frac{1}{3} b^mb_m
     +\frac{1}{3} M\BM
     +i\e am\tD_m b^a 
     -\bfD~,
\label{z78}
\ena
reproduces exactly the lagrangian (\ref{z66}) of curvature squared $U(1)$ supergravity.

On the other hand, performing appropriately a Weyl rescaling in the supersymmetric context \cite{WeB83}, \cite{HOW96f}, this action will describe supergravity, with a properly normalized curvature scalar, coupled to two chiral matter multiplets in the presence of a Fayet--Iliopoulos term.

%\mynewpage
%
%************
% CONCLUSION
%************

%\section{Conclusion}

In conclusion, we expect that this mechanism might open new possibilities for scenarios of gravity induced supersymmetry breaking in the presence of curvature squared terms \cite{HOW96f}.

%
%*****************
% ACKNOWLEDGMENTS
%*****************

\section{Acknowledgements}

I would like to thank R.Grimm for suggestions and helpful discussions about this work. Thanks also to G.Girardi for
conversations on this subject.

\mynewpage

%*************************************************************************
%                BIBLIOGRAPHIE
%*************************************************************************

\def\bibname{References}
%\addcontentsline{toc}{section}{Bibliographie}
\addcontentsline{toc}{section}{\bibname}

%\ {alpha} {plain} {unsrt} {abbrv}acm 
%% bibliographystyle{plb}
%% bibliography{i/cc,i/extended,i/mybib,i/sugar,i/sbgc,i/chern}

%********************************************************************************

\end{document}